\pdfoutput=1

\documentclass{jcgt}

\setciteauthor{Vinícius da Silva, Tiago Novello, Hélio Lopes, Luiz Velho}
\setcitetitle{Proceduray -- A light-weight engine for procedural primitive ray tracing.}

\submitted{\today}

\begin{document}

%==== Begin of code listing setup ====%

\definecolor{PigBlue}{RGB}{42, 0, 255}
\definecolor{PigRed}{RGB}{255, 0, 0}

\lstdefinelanguage{cpp}{
    language=C++,
    keywords=[1]{build, buildRays, buildHitGroups, buildGeometry, buildConstantBuffers, buildInstanceBuffer, buildAccelerationStructure, buildRootSignatures, buildShaderTablesEntries, addRay, make_shared, Payload, RayPayload, ShadowRayPayload, addHitGroup, addGeometry, buildInstancedProcedural, buildInstancedParallelepipeds, getD3DDevice, getBackBufferCount, getGeometry, getType, getInstances, create, buildGlobalRootSignature, buildLocalRootSignatures, createRange, getIndexBuffer, getGeometryMap, addDescriptorTable, addEntry, RootComponent, DontApply, getBuilded, SceneConstantBuffer, InstanceBuffer, addConstant, PrimitiveConstantBuffer, setRootArgumentsType, RootArguments, TriangleRootArguments, PrimitiveInstanceConstantBuffer, ProceduralRootArguments, addLocalSignature},
    keywordstyle=[1]\color{PigBlue},
    keywords=[2]{void, auto, for, if},
    keywordstyle=[2]\color{PigRed},
    morestring=[b]"
}

%==== End of code listing setup ====%

\title{Proceduray -- A light-weight engine for procedural primitive ray tracing.}
\author
    {Vinícius da Silva\\PUC-Rio
    \and Tiago Novello\\Visgraf-IMPA
    \and Hélio Lopes\\PUC-Rio
    \and Luiz Velho\\Visgraf-IMPA}

\teaser{
  \includegraphics[width=\textwidth]{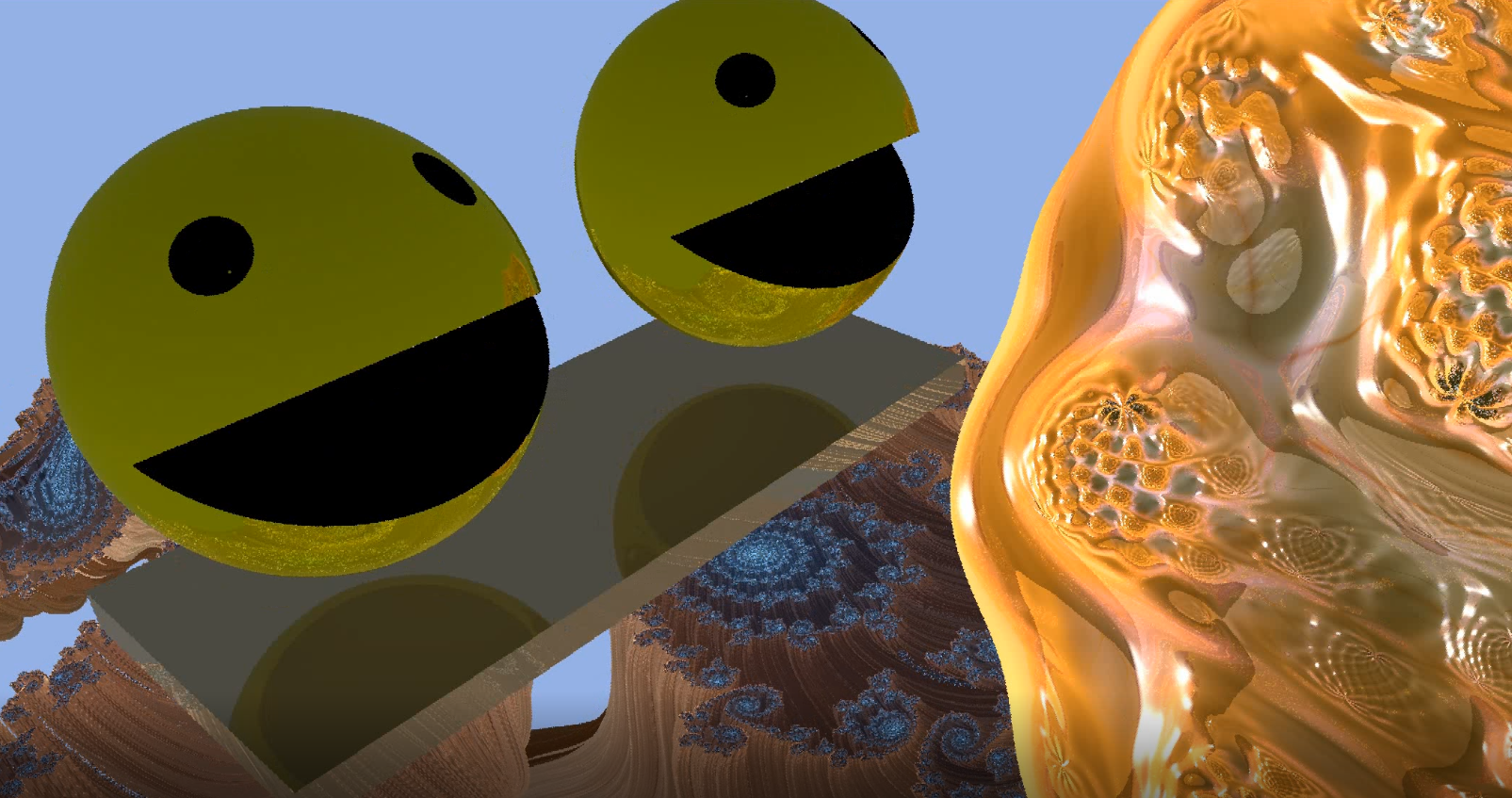}
  \caption{A scene with procedural objects (two pac-men, a julia set and a mandelbulb).}
  \label{fig:teaser}
}

\maketitle

\abstract{\small We introduce Proceduray, an engine for real-time ray tracing of procedural geometry. Its motivation is the current lack of mid-level abstraction tools for scenes with primitives involving intersection shaders. Those scenes impose strict engine design choices since they need flexibility in the shader table setup. Proceduray aims at providing a fair tradeoff between that flexibility and productivity. It also aims to be didactic. Shader table behavior can be confusing because parameters for indexing come from different parts of a system, involving both host and device code. This is different in essence from ray tracing triangle meshes (which must use a built-in intersection shader for all objects) or rendering with the traditional graphics or compute pipelines. Additionals goals of the project include fomenting deeper discussions about DirectX RayTracing (\textit{DXR}) host code and providing a good starting point for developers trying to deal with procedural geometry using DXR.}

\section{Introduction}
\label{sec:intro}

Given the potential of the RTX architecture for real-time graphics applications, important game engines~\cite{haas2014history,unrealengine}, scientific graphics frameworks~\cite{Benty20} and 3D creation software~\cite{Blender18} hastily incorporated it. Currently, several important 3D development tools support ray tracing triangle geometry using RTX.

\subsection{RTX in 3D development tools}

Tools supporting high-level abstraction workflows using RTX are limited to Falcor~\cite{Benty20}, Unreal~\cite{unrealengine}, and Unity~\cite{haas2014history}. There are other choices for development using RTX, but they are either lower-level abstraction libraries~\cite{mcguire2019introduction,parker2010optix,sellers2016vulkan,Haines2019} or non-interactive ray tracers~\cite{Mours19}.

Those tools differ in how they approach RTX, however. For example, Falcor~\cite{Benty20} has a more straightforward integration with the platform, providing very informative samples, including code to use ray-generation, closest-hit, any-hit, and miss shaders, and a path tracer. It did not support intersection shaders or procedural geometry until version 4.3, which debuted in December 2020. This version documents its API for working with procedural geometry, but it does not have examples on the matter yet.

Unity~\cite{haas2014history} and Unreal~\cite{unrealengine} have specific development branches \cite{Unity20,NVidia20} with RTX enabled. As Falcor, Unity supports customized RTX shaders~\cite{Ning19,NingPresentation19}, but does not work with intersection shaders. Unreal takes a different approach. Even though it has the most sophisticated RTX-based real-time ray tracer of all the alternatives, it can only be used as an effect, which can be turned on or off and be given parameters. We conjecture that Unreal has everything needed to support customized RTX shaders internally, but it currently lacks documentation and examples on the matter~\cite{Forum1,Forum2,Forum3}.

\subsection{Real-time ray-traced procedural geometry}

The fact that more than two years after the RTX launch the majority of the state-of-the-art 3D development tools do not support it in its full potential evidence that deeper discussions on the matter are necessary.

Although triangle geometry is most common for rendering purposes, RTX also supports procedural geometry. This feature includes control of the intersection behavior, which directly impacts how expressive a ray type can be. An immediate consequence is another level of flexibility available to applications. Additionally, procedural geometry imposes little resource maintenance in comparison with triangles since device code is directly responsible for defining geometry, instead of just receiving it to process. A discussion that emerges from those points is how to deal with the procedural geometry inherent flexibility in a productive workflow supporting high-level abstractions.

\subsection{Contributions and proposal}

We propose Proceduray, a novel lightweight engine with native support for procedural primitive ray tracing, designed to be a fair compromise between flexibility and productivity. Other objectives of the project include expanding the discussion about DXR~\cite{Dxr20} host code and providing a good starting reference for procedural geometry in DXR. Currently, references about DXR device shader code are abundant~\cite{Haines2019,Silva19}, but there are very scarce references about host code management, usually reduced to code samples in low-level abstraction libraries.

Proceduray was used to create all code for the chapter Real-time Rendering of Complex Fractals~\cite{dasilva2021fractals}, from Ray Tracing Gems 2~\cite{marrs2021ray}. This paper documents all host code for that chapter.

It is structured as follows. Section~\ref{sec:background} presents background DXR concepts, necessary to understand the problems and the discussion. Section~\ref{sec:design} describes in detail the problems that Proceduray deals with and the design choices to solve them. Section~\ref{sec:host} discusses the host API. Finally, Section~\ref{sec:conclusion} concludes and discusses future works.

\section{Background}
\label{sec:background}

DXR introduced a new rendering pipeline and several new graphics concepts with it. Since our objective is to write an engine using this technology, it is natural to first understand those ideas and how they relate with well-established ones, previously introduced by the rasterization and compute pipelines. We opt to be explicit and self-contained for the sake of novice developers. That said, most of this Section is an overview of very basic graphics development concepts and can be skipped by more experienced practitioners.

\subsection{Summary}

At a high abstraction level, the host application must perform a few tasks before being able to dispatch ray tracing using DXR:

\begin{itemize}
    \item Define an efficient way to detect ray-geometry intersections;
    \item Customize behavior based on geometry type. In other words, a specific geometry should run a specific set of shaders;
    \item Specify the resources needed for the shaders.
\end{itemize}

Our approach is to organize the background dependency concepts and develop from there until we know in detail how to perform each task, focusing on practical host resource management.

\subsection{Graphics Concepts}

Our objective in this section is to understand the building blocks needed to make resources available to DXR shaders, their dependencies, and their relationships.

\subsubsection{Overview}

\begin{figure}[!h]
    \centering
    \includegraphics[width=0.8\textwidth]{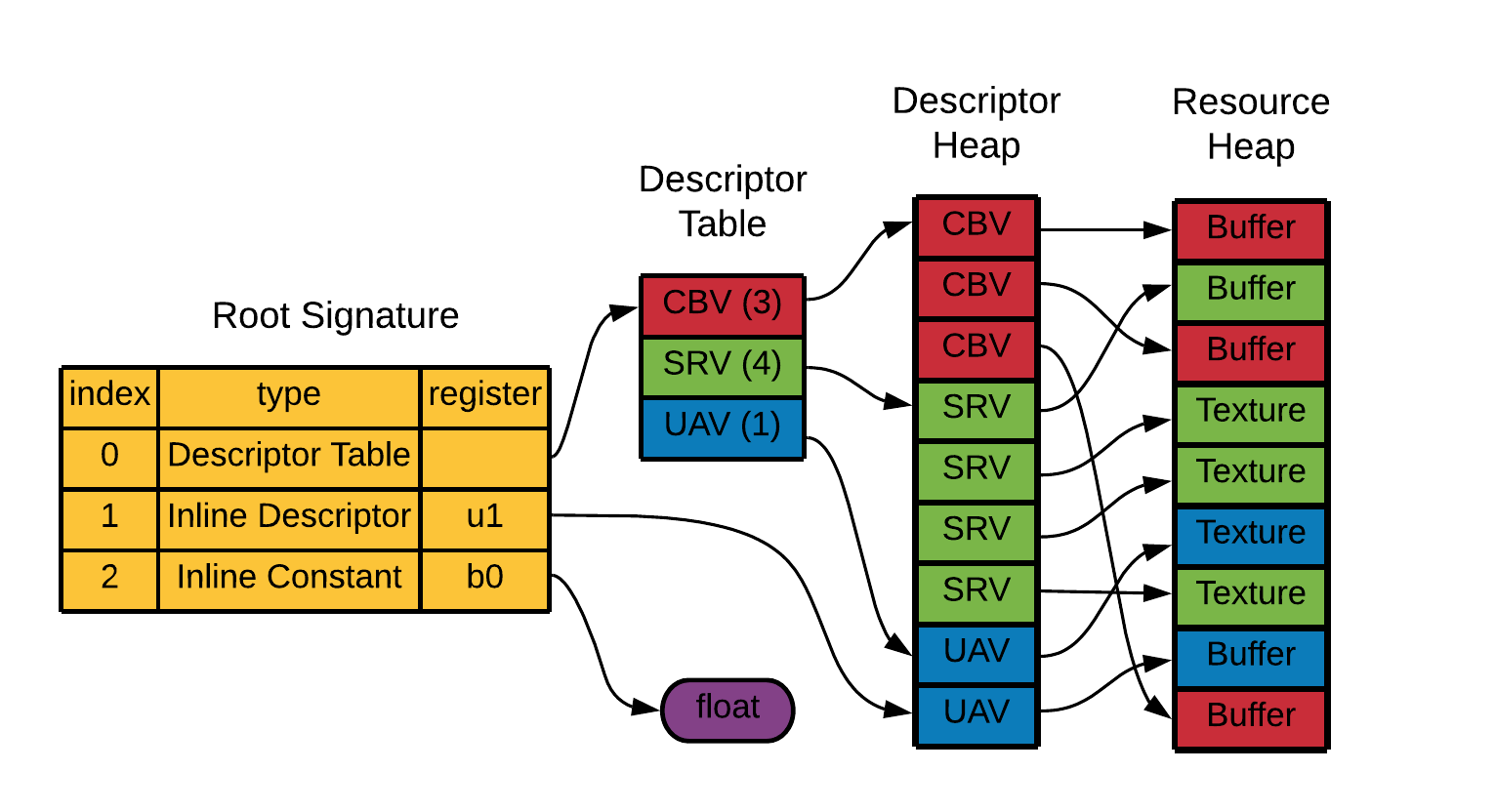}
    \vspace{-0.3cm}
    \caption{Overview of a Root Signature. Its entries can be Descriptor Tables, Inline Descriptors, or Inline Constants. Read the next Sections for more details.}
    \label{fig:root_signature}
\end{figure}

Every HLSL shader has an associated \textit{Root Signature}, which is one of the most important concepts for shader resource management. The term Signature is not chosen randomly here: a shader Root Signature is analogous to a programming language function signature. As a function signature describes the arguments needed for a function, a Root Signature describes the resources needed for a shader. Figure~\ref{fig:root_signature} contains an overview of a Root Signature. We will explain its components in detail in the next sections.

For didactic purposes, the concepts are organized into two groups: graphics pipeline and ray tracing pipeline. In practice, however, they are mixed in applications.

\subsubsection{Graphics pipeline}

\paragraph{Resource:}

It is every non-thread-local memory that is referenced by a shader, such as textures, constant buffers, images, and so on. Differently from thread-local variables, which are stored directly in device registers, \textit{Resource} registers contain indirections for the actual data in a Resource Heap. Figure~\ref{fig:root_signature} depicts such an example, where register $u1$ is an indirection to a Buffer. To make a resource available at shader execution time, one needs to use the concepts of \textit{Root Signatures}, \textit{Descriptors}, and \textit{Views}.

%\paragraph{Binding:}

%It is the process of resolving the indirection to access a \textit{Resource}, so it is available to the shader at execution time.

\paragraph{Descriptor:}

\textit{Resources} are not bind directly to the shader pipeline; instead, they are referenced through a \textit{Descriptor}. A \textit{Descriptor} is a small object that contains information about one resource. It resides in a \textit{Descriptor Heap} (see Figure~\ref{fig:root_signature}).

\paragraph{Descriptor Heap:}

a container for \textit{Descriptors}.

\paragraph{Descriptor Table:}

a set of references to \textit{Descriptors} in \textit{Descriptor Heaps}. It is composed of \textit{Descriptor Ranges}. In Figure~\ref{fig:root_signature}, entry 0 in the \textit{Root Signature} is a \textit{Descriptor Table}.

\paragraph{Descriptor Range:}

a range of consecutive \textit{Descriptors} in a \textit{Descriptor Heap}. In Figure~\ref{fig:root_signature}, the \textit{Descriptor Table} entry is composed of 3 \textit{Descriptor Ranges}, one for each \textit{View} type. In the example all ranges are defined in the same \textit{Descriptor Heap} for simplicity, but this is not an imposed restriction.

\paragraph{Views:}

\textit{Resources} are raw memory and \textit{Views} describe how they can be interpreted. The most common types are:

\begin{enumerate}
    \item \textit{Constant Buffer Views} (\textit{CBVs}): structured buffers. In practice, they are structs transferred from host to device code.
    \item \textit{Shader Resource Views} (\textit{SRVs}): typically wrap textures in a format that shaders can access them.
    \item \textit{Unordered Access Views} (\textit{UAVs}): enable the reading and writing to the texture (or other resources) in any order. The other types just support reading or writing, not both.
    \item Samplers: encapsulate sampling information for a texture, such as a filter, uv-coordinate, and level-of-detail parameters.
\end{enumerate}

\paragraph{Root Signatures:}

Before DXR, only \textit{Global Root Signatures} existed: when one was bound, it was visible to all shaders dispatched. Since DXR supports per-geometry customized shaders, \textit{Root Signatures} can also be Local now. \textit{Global Root Signatures} continue to be visible to all shaders, but \textit{Local Root Signatures} can be set up to be used only when a specific geometry is ray-traced.

Figure~\ref{fig:root_signature} shows an example of a \textit{Root Signature}, which can be composed by:

\begin{enumerate}
    \item \textit{Inline Constants}, which are structs inlined directly at the \textit{Root Signature};
    \item \textit{Inline Descriptors}, which are \textit{Descriptors} inlined directly at the \textit{Root Signature};
    \item \textit{Descriptor Tables}, which are references to \textit{Descriptors} in \textit{Descriptor Heaps}.
\end{enumerate}

\subsubsection{Ray tracing pipeline}

\paragraph{Ray:}

the entity that will be used to ray trace. It is composed of a \textit{Payload}, which accompanies it throughout all the pipeline. This \textit{Payload} is customizable (in practice, can be any plain-old-data (POD) struct) and can be updated by the ray tracing shaders.

\begin{figure}[!h]
    \centering
    \includegraphics[width=0.7\textwidth]{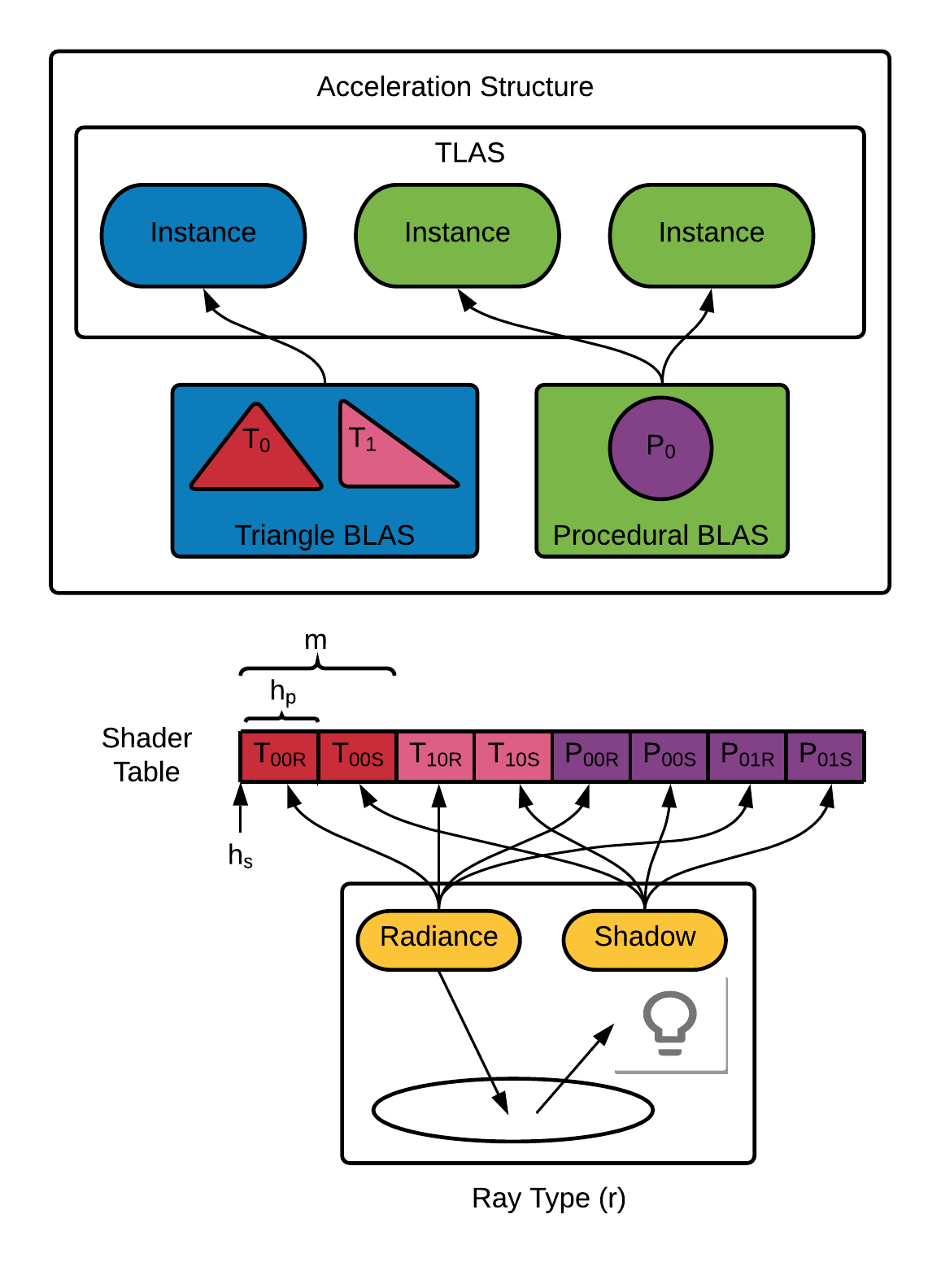}
    \vspace{-0.6cm}
    \caption{Overview of an \textit{Acceleration Structure}, with associated \textit{Shader Table}. In the example, it is composed of two \textit{Bottom Level Acceleration Structures} (\textit{BLAS}), which are instanced in the \textit{Top Level Acceleration Structure} (\textit{TLAS}). Notice how the Triangle \textit{BLAS} is composed of two geometries. What happens when a node is reached in a traversal depends on the \textit{Shader Table Indexing Rule}. For more details, continue reading this Section.}
    \label{fig:shader_table}
\end{figure}

\paragraph{Acceleration Structure:}

a hierarchical spatial data structure used to accelerate ray-geometry intersections. It is composed of \textit{Bottom-Level Acceleration Structures (BLAS)} and \textit{Top-Level Acceleration Structures (TLAS)}. \textit{BLAS} are the acceleration structures where the geometry lies in and \textit{TLAS} are used for instancing (reusing) \textit{BLAS}. The Acceleration Structure creation is performed by the DXR runtime but can be customized by parameters. \textit{BLAS} creation for triangle and procedural geometry are different. A  data structure optimized for triangle culling is constructed for the triangle case, while a simpler structure enclosing procedural geometry axis-aligned bounding boxes (AABBs) is constructed for the procedural case. Customized \textit{Intersection Shaders} are responsible for the remaining intersection tests inside the AABBs. Figure~\ref{fig:shader_table} shows an overview of \textit{Acceleration Structures} and \textit{Shader Tables}, which will be detailed later on.

\paragraph{Ray Tracing Shaders:}

They are called at different moments of the \textit{Acceleration Structure} traversal by the DXR runtime.

\begin{enumerate}
    \item \textit{Ray Generation Shader}: the starting point, where all initial rays are launched. Usually, each ray starts at the camera position and goes through a pixel.
    \item \textit{Miss Shader}: issued when a ray misses all geometry. Receives the ray \textit{Payload} as input.
    \item \textit{Any-hit Shader}: called for all primitive intersections found. Its inputs are the ray \textit{Payload} and a structure coming from the \textit{Intersection Shader}.
    \item \textit{Closest-hit Shader}: called for the primitive intersection closest to the ray origin. It has the same input types as the \textit{Any-hit Shader}.
    \item \textit{Intersection Shader}: called for computing intersections inside a \textit{BLAS} AABB node. Custom intersection shaders are specific to procedural geometry (a built-in optimized \textit{Intersection Shader} is used for triangle geometry). The result of the intersection shader indicates if an intersection is found, and the data that must be fed to the \textit{any-hit} and \textit{closest-hit shaders} potentially called by the runtime in consequence.
\end{enumerate}

\paragraph{Hit Group:}

a set of shaders that deal with a specific geometry. Specifically, one or more shaders consisting of: 0 or 1 \textit{Intersection Shader}, 0 or 1 \textit{Any-hit Shader}, 0 or 1 \textit{Closest-hit Shader}. Hit groups are used in conjunction with \textit{Shader Tables} and the \textit{Shader Table Indexing Rule} to enable customized behavior for geometries. \textit{Ray Generation} and \textit{Miss Shaders} cannot be part of a \textit{Hit Group} because they aren’t involved directly with geometry.

\paragraph{Shader Tables:}

When traversing the \textit{Acceleration Structure}, the runtime needs to know which \textit{Hit Group} and shader parameters must be used when a \textit{BLAS} node is reached. This information is fed to the pipeline through \textit{Shader Tables}.

A \textit{Shader Table} is composed of \textit{Ray-generation Shader Tables}, \textit{Miss Shader Tables} and \textit{Hit-group Shader Tables}. We are interested on the \textit{hit-group shader} tables, which enable customized behavior for geometries. They are very flexible and their entries must be set according to the \textit{Indexing Rule}. This is the most important (and confusing) concept regarding \textit{Acceleration Structure} behavior and flexibility. It is defined by the following equation:
\begin{equation}
\label{eq:indexing}
    a = h_s + h_p * (r + m * g + i)\cite{Dxr20},
\end{equation}

\noindent where:

\begin{itemize}
    \item $a$ is the address of the \textit{Shader Table} record;
    \item $h_s$ is the \textit{Hit Group Shader Table} start address;
    \item $h_p$ is the \textit{Hit Group Shader Table} stride;
    \item $r$ is the ray type contribution;
    \item $m$ is a multiplier for the geometry contribution;
    \item $g$ is the geometry contribution;
    \item $i$ is the instance contribution.
\end{itemize}

Figure~\ref{fig:shader_table} shows an example. There, $h_s$ is the pointer to the beginning of the table. $h_p$ is the size of each entry. That example has two \textit{Ray} types: Radiance and Shadow. $m$ is set to 2 to reflect this fact. In other words, each geometry has an entry for when it is hit by Radiance \textit{Rays} and another for Shadow \textit{Rays}. With this setup, a \textit{Hit Group} can be used for each case. This is expected since the shadow query is a much simpler operation. $g$ is an ordered index automatically set by the DXR runtime to identify a geometry inside a \textit{BLAS}. In this example, $T_0$ has index 0 and $T_1$ has index 1. Finally, $i$ is set by the application when creating the $TLAS$, so the instances can be taken into consideration when indexing. In the example, the Procedural \textit{BLAS} has two instances. If we want each one to be treated differently (which is the case), we set $i = 0$ for the first instance and $i = 1$ for the second instance. To reflect the \textit{Indexing Rule}, each entry in the \textit{Shader Table} of Figure~\ref{fig:shader_table} has the form $G_{gi[R|S]}$, where $G$ indicates the Geometry type ($T$ or $P$), $g$ and $i$ are the parameters in Equation~\ref{eq:indexing} and $R$ or $S$ are possible values for $r$, indicating the \textit{Ray} type ($R$ for Radiance and $S$ for Shadow). The contents of the entries are defined by the application, but they usually contain data to fetch buffers in the \textit{Global Root Signature}. It is important to note that Equation~\ref{eq:indexing} is very flexible and there is no unique way to use the parameters. A central design choice is how the application deals with the \textit{Indexing Rule}.

One additional property makes Equation~\ref{eq:indexing} a little bit more confusing. Its parameters come from elements residing both at the host and device code, and each one has a specific way to be set. This is also the main reason why DXR concepts cannot be completely decoupled and should always be thought about regarding their global meaning.

\begin{itemize}
    \item $h_s$ and $h_p$ must be set by the host code when dispatching the ray tracing device code by calling \verb|DispatchRays()|;
    \item $r$ and $m$ must be set by HLSL ray tracing shaders when calling \verb|TraceRay()|, which is the intrinsic responsible for launching specific rays;
    \item $g$ is a sequential index automatically generated by the runtime at \textit{BLAS} creation.
    \item $i$ is set by the host code when creating the \textit{TLAS}.
\end{itemize}

\paragraph{Ray Tracing Pipeline State Object (RTPSO):}

\textit{Shader Tables} associate \textit{Acceleration Structures} with \textit{Hit Groups}. However, \textit{Hit Groups} also need an associated \textit{Root Signature} so the \textit{Resources} needed by the shaders are known at execution time. This association is performed by the \textit{RTPSO}, which represents a full set of shaders that could be reached by a dispatch call, with all configuration options and associations resolved. It is composed of \textit{Subobjects}, including:

\begin{itemize}
    \item \textit{Shader Library (DXIL) Subobject}: contains the compiled shaders;
    \item \textit{Hit Groups Subobjects}: define the shader entry points;
    \item \textit{Root Signature Subobjects}: contains \textit{Root Signatures};
    \item \textit{Shader Export Association Subjects}: associate \textit{Hit Group Subobjects} with \textit{Local Root Signature Subobjects}.
\end{itemize}

\subsection{Summary revisited}

To wrap up, we revisit and update the Summary, given that we now know all the concepts needed to set up ray tracing. The updated tasks before being able to dispatch ray tracing are:

\begin{itemize}
    \item Creating an \textit{Acceleration Structure} to detect intersections with the geometry.
    \item Defining a \textit{Shader Table} to indicate which \textit{Hit Group} and shader parameters must be used when ray tracing a given geometry;
    \item Creating a \textit{RTPSO} to indicate which \textit{Local Root Signature} should be used for a given geometry, which defines all \textit{Resources} needed by the shaders in the \textit{Hit Group}.
\end{itemize}

\section{Proceduray: Design Choices}
\label{sec:design}

As could be seen in Section~\ref{sec:background}, the concepts needed to perform real-time ray tracing using DXR are highly coupled, in special \textit{BLAS}, \textit{Hit Groups}, \textit{Shader Tables}, \textit{Local Root Signatures} and \textit{RTPSOs}. On one hand, we want Proceduray to be flexible to support a variety of procedural geometry applications. On the other hand, too much flexibility can harm the capacity of the engine to provide high-level abstractions that can increase productivity. The design choices to address those and other difficulties are described next.

\subsection{Design choices}

\subsubsection{New entity types}

Since the association between concepts can be intricate, the first design choice is to support several new entity types in scenes, so the user can be flexible in the associations needed by the application. Proceduray supports \textit{Ray}, \textit{Global Root Signature}, \textit{Local Root Signature} and \textit{Hit Group} entities in addition to the usual Geometry entities.

\subsubsection{Shader Compatibility Layer}

One of the most common problems when writing shaders is to ensure that data transfer between host and device code is robust. Since there is an entire pipeline between the host code and the final image, a lot of time can be lost debugging errors in this area. A usual approach in engines is to fix a compatibility layer for the effects supported by the engine.

To work with procedural shaders, this compatibility layer must be more flexible.  Proceduray has a shader compatibility component that centralizes all POD structs that must be seen in host and device code, but the user defines all of them. The only prerequisite is to classify the structs in predefined categories: \textit{Payloads}, \textit{Root Components}, \textit{Root Arguments} and \textit{Attribute Structs}.

\begin{enumerate}
    \item \textit{Payloads} are the types used for ray payloads in the shaders.
    \item \textit{Root Components} are any type that can be sent in a Root Signature as \textit{Root Constants} or \textit{Root Descriptors}.
    \item \textit{Root Arguments} are structs with the actual values that a shader associated with the \textit{Root Signature} will see at execution time.
    \item \textit{Attribute Structs} are the types that \textit{intersection shaders} will fill when detecting an intersection so the \textit{any-hit} and \textit{closest-hit shaders} receive data to work with.
\end{enumerate}

This design helps the engine to automatically set several parameters needed for the DXR runtime, such as maximum payload, root argument, and attribute struct sizes. It is also the basis for designing Proceduray Type-safe Resources.

\subsubsection{Type-safe Resources}

Another design choice to ensure security in transfers between host and device code is to ensure type safety for all types declared by the user in the \textit{Shader Compatibility Layer}. Proceduray uses C$++$ $17$ variants to treat all types declared in a single category as a single type engine-wide.

\subsubsection{Scene entity queries}

Proceduray provides ids for every entity in a scene and fast queries by id using maps. Ids and queries are the core tools to associate entities for \textit{Shader Table} entries or \textit{RTPSO} Subobjects.

\subsubsection{BLAS from Geometry}

We choose to create a \textit{BLAS} automatically for each \textit{Geometry} entity, which can be marked as a triangle or procedural. Triangle geometry has the usual vertices and indices and Procedural geometry has an AABB only. The instances are defined by the application, at Geometry creation time. The resulting \textit{Acceleration Structure} has a \textit{BLAS} for each \textit{Geometry} and a \textit{TLAS} composed of all instances. The parameter $i$ in the \textit{Indexing Rule} (Eq.~\ref{eq:indexing}) is a sequential index depending on the order of instance creation.

\subsubsection{Shader Table entries decoupled from Shader Tables}

To increase flexibility, Proceduray creates \textit{Shader Tables} in two passes. The first is the definition of the table entries, which are created by the user. The second is a building pass that effectively creates the tables for the runtime. Decoupling the entries from the actual table helps to deal with the tight coupling between \textit{Shader Tables} and \textit{RTPSOs}.

\subsubsection{Shader table entries as tuples from scene entities}

Since we want users to create shader table entries, we facilitate the process by letting the entries be defined as tuples of entity ids and a few additional data. An entry tuple has the form \textit{$<$Ray id, Hit Group id, Local Root Signature id, Root Arguments$>$}. The first three parameters are ids to the entities in the scene and the last one is the \textit{Root Arguments} struct. The \textit{Root Arguments} struct contains the actual values that will be received by the shader at execution time. It is important to note that the type safety of the \textit{Root Arguments} struct is automatically ensured by the \textit{Shader Compatibility Layer}.

\subsubsection{RTPSO shader export association subobjects from Shader Table entries}

The \textit{Shader Table} entries are also used to automatically create the association between \textit{Hit Groups} and \textit{Local Root Signatures} in the RTPSO.

\section{Proceduray: Host API}
\label{sec:host}

\begin{figure}[!h]
    \centering
    \includegraphics[width=\textwidth]{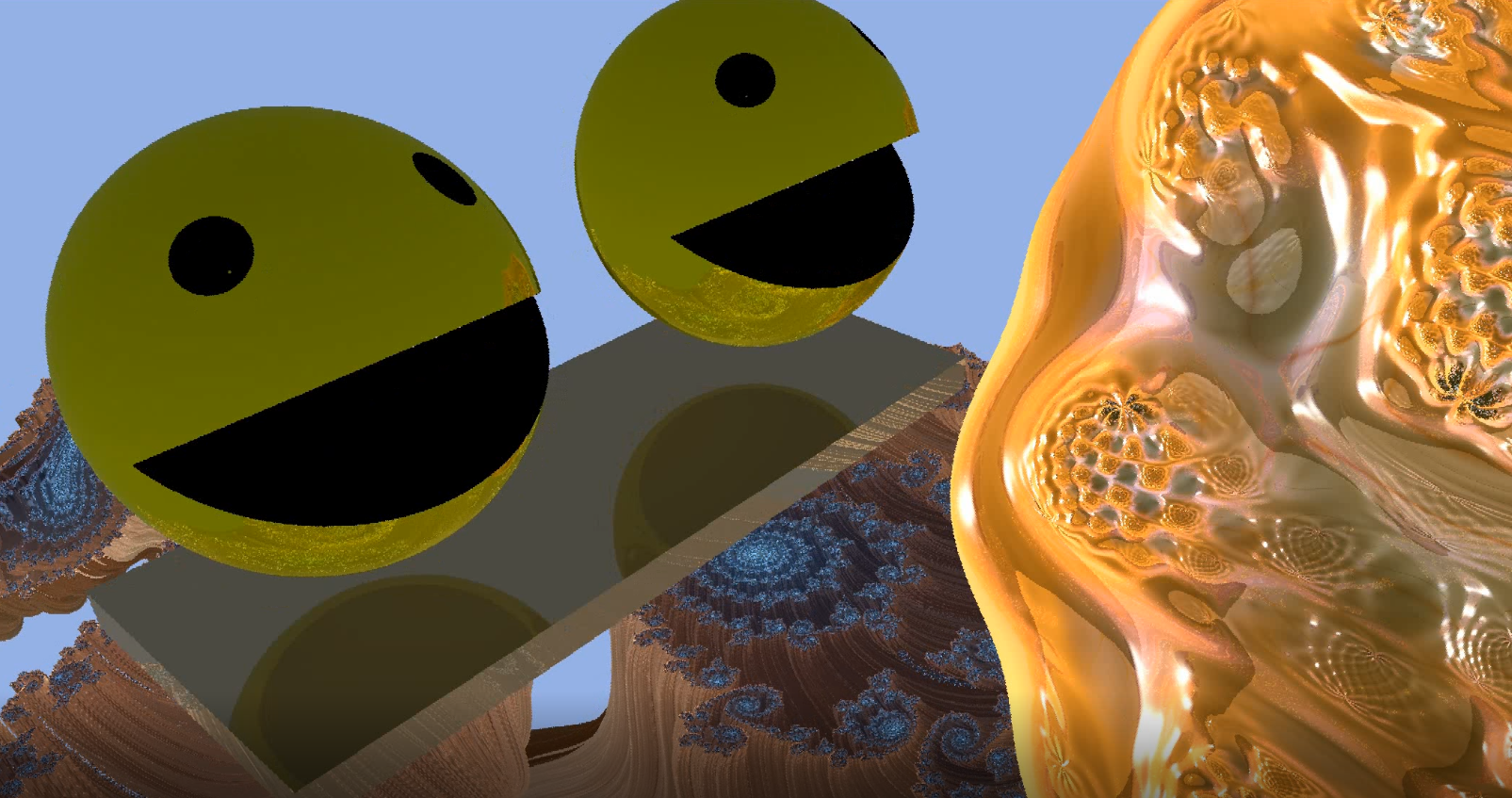}
    \vspace{-0.2cm}
    \caption{Sample scene: a triangle plane mesh and several procedural objects (two Pac-men, a 3D Julia Set and a Mandelbulb).}
    \label{fig:example_scene}
\end{figure}

\begin{figure}[!h]
    \centering
    \includegraphics[width=0.85\textwidth]{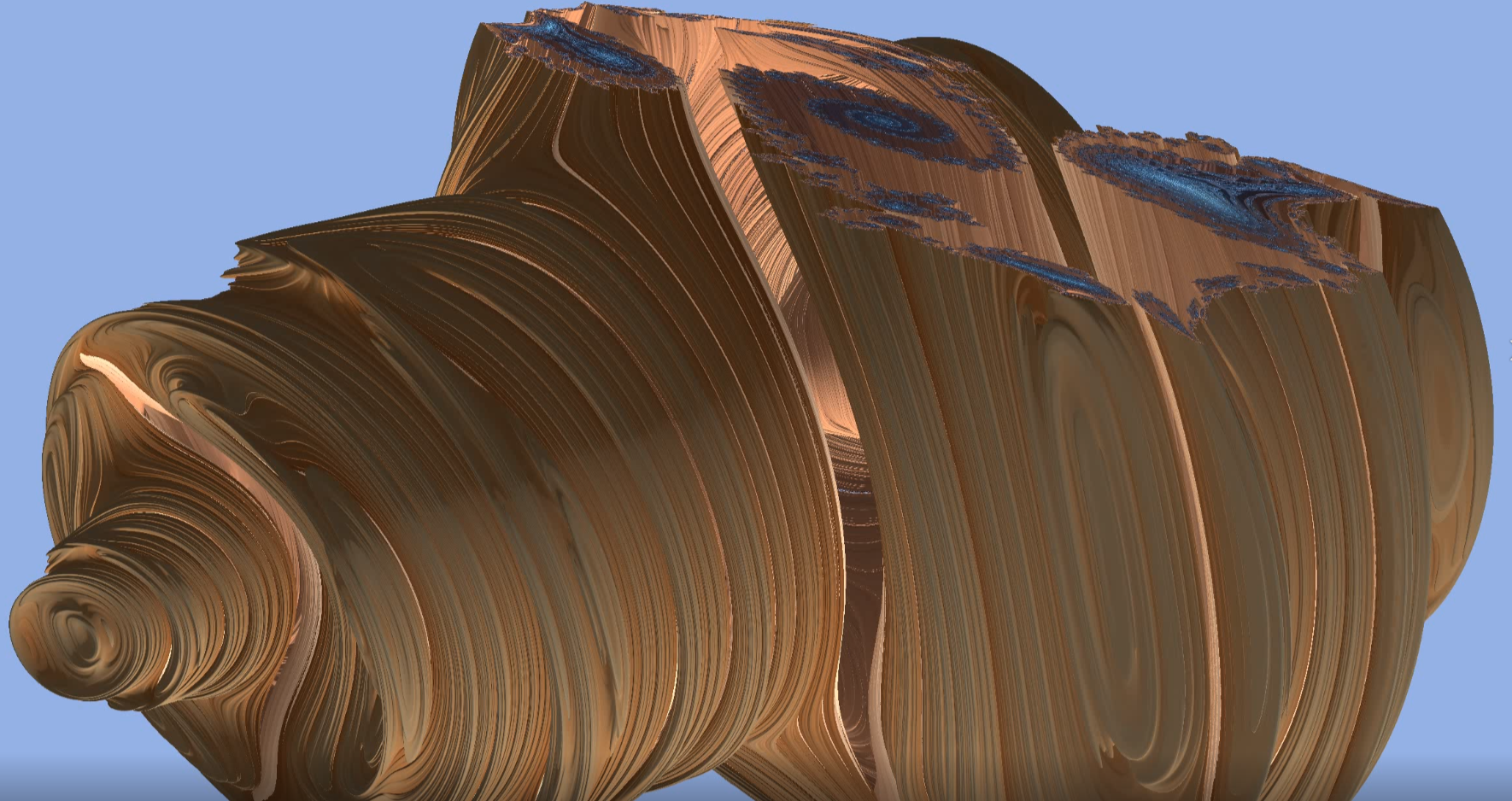}
    \includegraphics[width=0.85\textwidth]{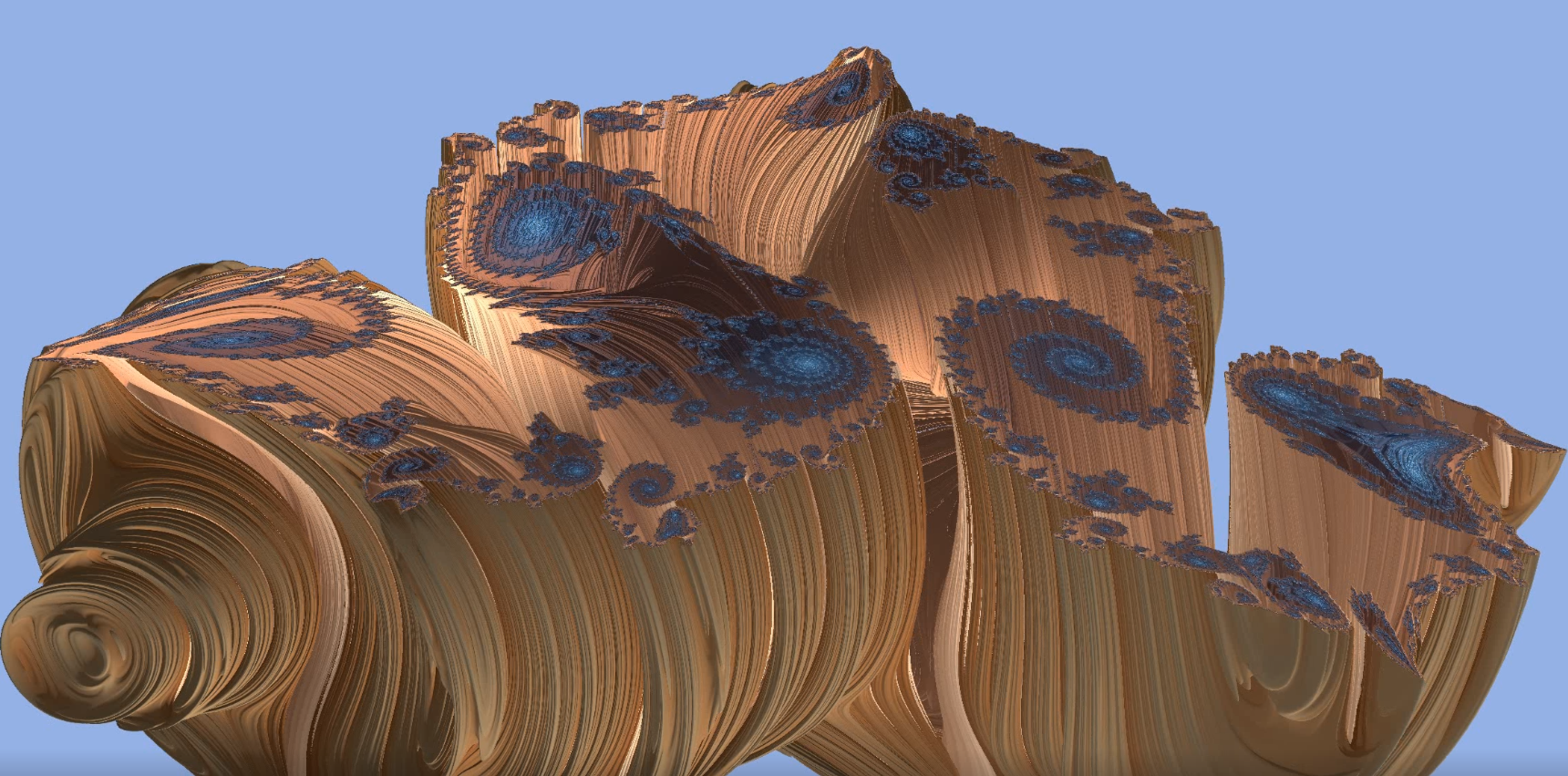}
    \caption{A 3D Julia set cut by two different planes.}
    \label{fig:cutJulia}
\end{figure}

\begin{figure}[!h]
    \centering
    \includegraphics[width=0.4\textwidth]{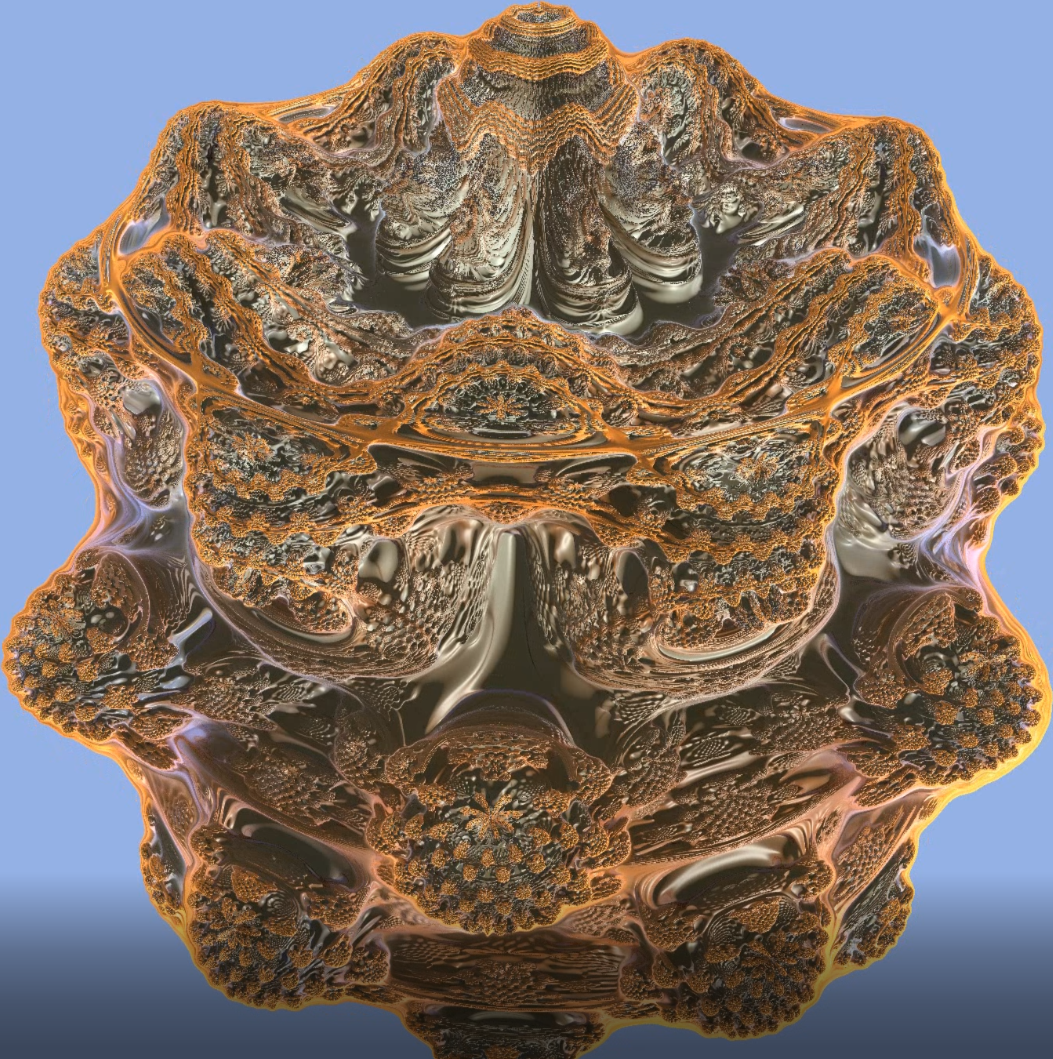}
    \includegraphics[width=0.45\textwidth]{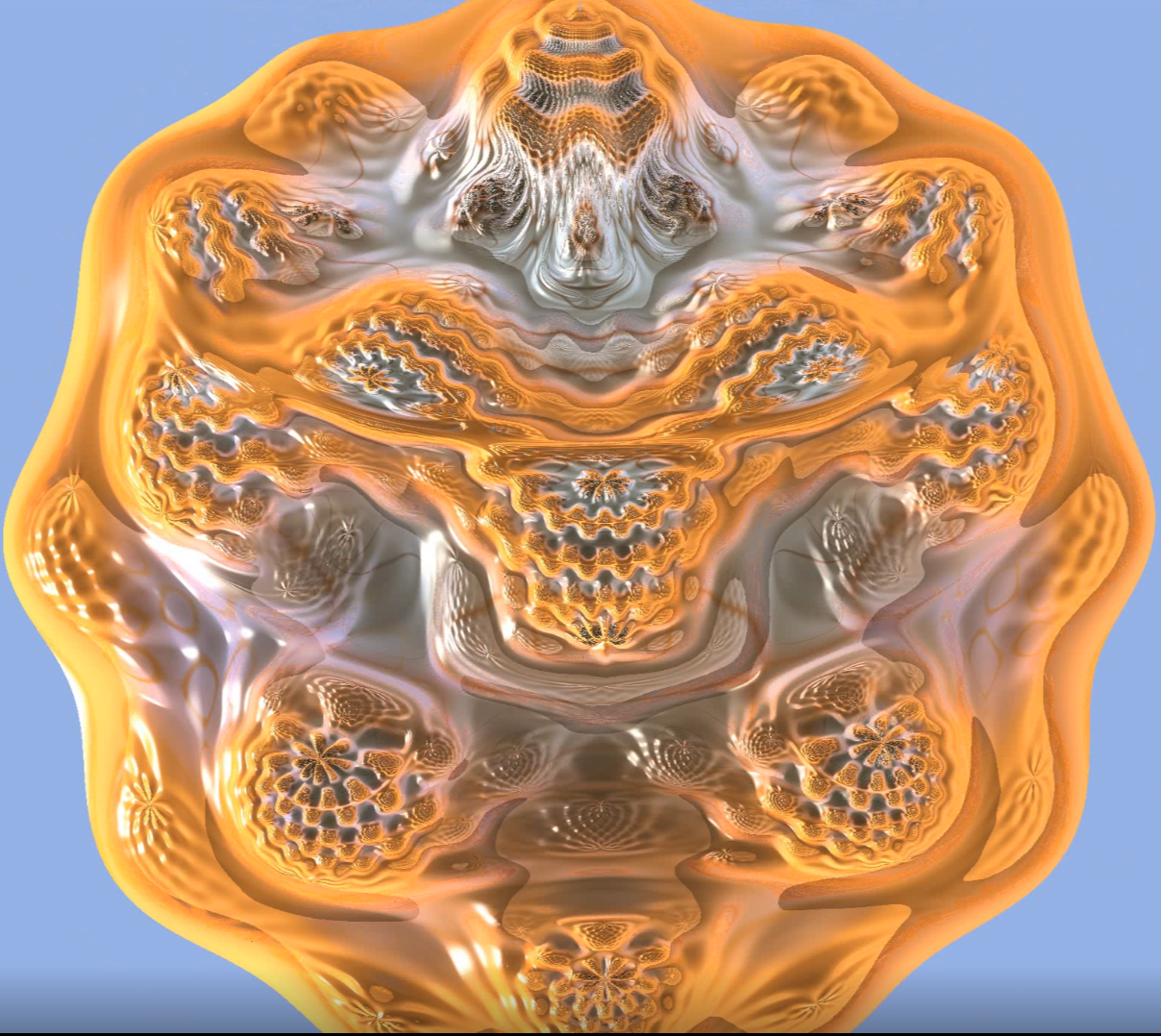}
    \caption{Two frames of an animated Mandelbulb. The animation is created varying the number of ray marching iterations over time.}
    \label{fig:mandelbulb}
\end{figure}

Proceduray's design choices reflect in its API, which is discussed in this section. Our approach is to show small simplified C++ code snippets to guide the creation of a sample scene containing a triangle plane mesh and several procedural objects (two CSG Pac-men, and two fractals: a 3D Julia Set, and a Mandelbulb). The full source code can be found in {\small \url{https://github.com/dsilvavinicius/Proceduray}}, implemented and tested on Windows, using Visual Studio 2019. For theory and details about the fractals shader code, please refer to~\cite{dasilva2021fractals}. Figure~\ref{fig:example_scene} shows the sample scene and Figures~\ref{fig:cutJulia} and \ref{fig:mandelbulb} show close-ups of a 3D Julia Set cut by different planes and an animated Mandelbulb, respectively.

\pagebreak

\subsection{Procedural Sample Scene}

The core of the sample is in the user class \lstinline[language=cpp]{MandelJuliaPacSceneBuilder}, which creates everything needed to set the \textit{RTPSO} up in function \lstinline[language=cpp]{build()} (Listing~\ref{code:build}). This includes \textit{Rays}, \textit{Hit Groups}, \textit{Geometry}, Constant Buffers for communication between host and device, an instance buffer with information fetched using the entries in the \textit{Shader Table} and the \textit{Indexing Rule} (Equation~\ref{eq:indexing}), the \textit{Acceleration Structure}, \textit{Root Signatures} and \textit{Shader Table} entries.

\begin{lstlisting}[caption={The core sample function: MandelJuliaPacSceneBuilder::build().}, label={code:build}, language=cpp]
void MandelJuliaPacSceneBuilder::build()
{
	buildRays();
	buildHitGroups();
	buildGeometry();
	buildConstantBuffers();
	buildInstanceBuffer();
	buildAccelerationStructure();
	buildRootSignatures();
	buildShaderTablesEntries();
}
\end{lstlisting}

Each one of the methods called use Proceduray's API to create the objects needed. \lstinline[language=cpp]{buildRays()} creates all \textit{Ray} types and specifies string ids for future queries. This pattern is repeated for almost all scene components. The API expects the \textit{miss shader} entry point and the \textit{Payload}, which is a variant for the Shader Compatibility Layer. The sample has a radiance ray type and a shadow ray type (Listing~\ref{code:rays}).

\begin{lstlisting}[caption={Ray creation.}, label={code:rays}, language=cpp]
void MandelJuliaPacSceneBuilder::buildRays()
{
	m_scene->addRay(
	    make_shared<Ray>(
	        "Radiance", L"Miss", Payload(RayPayload())
        )
    );
	m_scene->addRay(
	    make_shared<Ray>(
	        "Shadow", L"Miss_Shadow", Payload(ShadowRayPayload())
        )
    );
}
\end{lstlisting}

The API expects a tuple \textit{$<$id, internal name, any-hit shader entry point, closest-hit shader entry point, intersection shader entry point$>$} to represent a \textit{Hit Group}, which is added to the scene using \lstinline[language=cpp]{addHitGroup()}. The entry points can be empty if the \textit{Hit Group} does not have the associated shaders. Function \lstinline[language=cpp]{buildHitGroups()} (Listing~\ref{code:hitgroups}) shows how to add hit groups for the radiance and shadow rays used in the sample. Notice how shadow \textit{Hit Groups} only needs to define the \textit{Intersection Shader} (except for the plane, which must use the built-in one).

\begin{lstlisting}[caption={Hit group creation.}, label={code:hitgroups}, language=cpp]

void MandelJuliaPacSceneBuilder::buildHitGroups()
{
	// Triangle Hit Groups.
	m_scene->addHitGroup(make_shared<HitGroup>(
	    "Triangle", L"HitGroup_Triangle", L"",
	    L"ClosestHit_Triangle", L""));
	
	m_scene->addHitGroup(make_shared<HitGroup>(
	    "Triangle_Shadow", L"HitGroup_Triangle_Shadow",
	    L"", L"", L""));

	// Procedural Hit Groups.
	m_scene->addHitGroup(make_shared<HitGroup>(
	    "Pacman", L"HitGroup_Pacman", L"",
	    L"ClosestHit_Pacman", L"Intersection_Pacman"));
	
	m_scene->addHitGroup(make_shared<HitGroup>(
	    "Pacman_Shadow", L"HitGroup_Pacman_Shadow",
	    L"", L"", L"Intersection_Pacman"));

	m_scene->addHitGroup(make_shared<HitGroup>(
	    "Mandelbulb", L"HitGroup_Mandelbulb", L"", 
	    L"ClosestHit_Mandelbulb", L"Intersection_Mandelbulb"));
	
	m_scene->addHitGroup(make_shared<HitGroup>(
	    "Mandelbulb_Shadow", L"HitGroup_Mandelbulb_Shadow",
	    L"", L"", L"Intersection_Mandelbulb"));

	m_scene->addHitGroup(make_shared<HitGroup>(
	    "Julia", L"HitGroup_Julia", L"",
	    L"ClosestHit_Julia", L"Intersection_Julia"));
	
	m_scene->addHitGroup(make_shared<HitGroup>(
	    "Julia_Shadow", L"HitGroup_Julia_Shadow",
	    L"", L"", L"Intersection_Julia"));
}

\end{lstlisting}

Geometry can be added to the scene using \lstinline[language=cpp]{addGeometry()}. Procedural geometry expects an AABB and an instance vector with transformations for proper instance placement in the scene. Triangle meshes expect arrays of vertices and indices, and an instance vector as well. Listings ~\ref{code:proc_geometry} and ~\ref{code:tri_geometry} show the code.

\pagebreak

\begin{lstlisting}[caption={Procedural geometry creation.}, label={code:proc_geometry}, language=cpp]

void MandelJuliaPacSceneBuilder::buildInstancedProcedural()
{
    auto deviceResources = m_dxr->deviceResources;
    
    Geometry::Instances mandelbulbInstances;
	Geometry::Instances pacManInstances;
	Geometry::Instances juliaInstances;
	
	// Create instance transformation matrices ...
    
    // Create geometry
	D3D12_RAYTRACING_AABB juliaAABB{ /* AABB position and size */ };
	m_scene->addGeometry(make_shared<Geometry>(
	    "Julia", juliaAABB, *deviceResources, juliaInstances));

	D3D12_RAYTRACING_AABB pacmanAABB{ /* AABB position and size */ };
	m_scene->addGeometry(make_shared<Geometry>(
	    "Pacman", pacmanAABB, *deviceResources, pacManInstances));

	D3D12_RAYTRACING_AABB mandelAABB{ /* AABB position and size */ };
	m_scene->addGeometry(make_shared<Geometry>(
	    "Mandelbulb",mandelAABB, *deviceResources, mandelInstances));
}

\end{lstlisting}

\begin{lstlisting}[caption={Triangle geometry creation.}, label={code:tri_geometry}, language=cpp]

void MandelJuliaPacSceneBuilder::buildInstancedParallelepipeds()
{
    vector<Vertex> vertices = // Load vertices.
    vector<Index> indices = // Load indices.
    
    m_scene->addGeometry(make_shared<Geometry>(
        "GlobalGeometry", vertices, indices, *m_dxr->deviceResources, 
        *m_dxr->descriptorHeap, instances));
}

\end{lstlisting}

The procedural shaders in the sample scene must access the transformation matrices so they can work at local space instead of world space. This approach simplifies the \textit{Intersection Shaders}. Those matrices are stored in a buffer, created by the function \lstinline[language=cpp]{buildInstanceBuffer()} (Listing~\ref{code:instance_buffer}).

\pagebreak

\begin{lstlisting}[caption={Instance buffer creation.}, label={code:instance_buffer}, language=cpp]

void MandelJuliaPacSceneBuilder::buildInstanceBuffer()
{
	auto deviceResources = m_dxr->deviceResources;
	auto device = deviceResources->getD3DDevice();
	auto frameCount = deviceResources->getBackBufferCount();

	UINT nProceduralInstances = 0;
	for (auto geometry : m_scene->getGeometry())
	{
		if (geometry->getType() == Geometry::Procedural)
		{
			for (auto instance : *geometry->getInstances())
			{
				++nProceduralInstances;
			}
		}
	}

	m_instanceBuffer->create(device, nProceduralInstances,
	    frameCount, L"AABB primitive attributes");
}

\end{lstlisting}

The Acceleration structure is created directly from scene geometry, as discussed in Section~\ref{sec:design}. Thus, the API for this is very straightforward, as can be seen in Listing~\ref{code:acceleration}.

\begin{lstlisting}[caption={Acceleration Structure creation.}, label={code:acceleration}, language=cpp]

void MandelJuliaPacSceneBuilder::buildAccelerationStructure()
{
	m_accelerationStruct = make_shared<AccelerationStructure>(
	    m_scene, m_dxr->device, m_dxr->commandList, 
	    m_dxr->deviceResources);
}

\end{lstlisting}

\subsubsection{Root Signatures and Shader Table Entries}

A key idea behind the sample is to express an efficient and flexible way to fetch geometry data using the \textit{Acceleration Structure}. Efficiency comes from minimizing data on the \textit{Local Root Signatures}, so fewer resources are needed for the intersection traversal and more parallel work can be done. This is done using the \textit{Local Root Signatures} data to index the actual geometry on the \textit{Global Root Signature}. In practical terms, the triangle vertex and index buffers, as well as the instance buffer are in the \textit{Global Root Signature}. The instance buffer is then indexed by the values contained in the \textit{Local Root Signatures}.

The \textit{Global Root Signature} contains the output texture, the acceleration structure, the scene constant buffer, the plane buffers, and the instance buffer. Each entry describes which register will be bound to which resource. Notice how the API needs a type specification for each inline entry. This type is ensured by the \textit{Shader Compatibility Layer} for all operations involving those components. Listing~\ref{code:global_root} shows the code.

\begin{lstlisting}[caption={Global Root Signature creation.}, label={code:global_root}, language=cpp]

void MandelJuliaPacSceneBuilder::buildGlobalRootSignature()
{
	auto deviceResources = m_dxr->deviceResources;
	
	// Global root signature.
	auto globalSignature = make_shared<RootSignature>(
	    "GlobalSignature", deviceResources, m_dxr->descriptorHeap,
	    false);

	// Global signature ranges.
	auto outputRange = globalSignature->createRange(
	    m_dxr->outputHandler.gpu, RootSignature::UAV, 0, 1);
	auto globalGeometry = m_scene->getGeometryMap()
	    .at("GlobalGeometry");
	auto vertexRange = globalSignature->createRange(
	    globalGeometry->getIndexBuffer().gpuDescriptorHandle, 
	    RootSignature::SRV, 1, 2);

	// Global signature entries.
	m_dxr->outputHandler.baseHandleIndex =
        globalSignature->addDescriptorTable(
        vector<RootSignature::DescriptorRange>{outputRange});
	globalSignature->addEntry(RootComponent(DontApply()),
	    RootSignature::SRV, m_accelerationStruct->getBuilded(), 0);
	globalSignature->addEntry(RootComponent(SceneConstantBuffer()), 
	    RootSignature::CBV, m_sceneCB, 0);
	globalSignature->addEntry(RootComponent(InstanceBuffer()),
	    RootSignature::SRV, m_instanceBuffer, 3);
	globalSignature->addDescriptorTable(
	    vector<RootSignature::DescriptorRange>{vertexRange});

	m_scene->addGlobalSignature(globalSignature);
}

\end{lstlisting}

The sample contains two \textit{Local Root Signatures}, one for triangles and another for procedural geometry (Listing~\ref{code:local_root}). Type safety is also applied for all types set in the process. The conjunction of the \textit{Shader Table} and the \textit{RTPSO} ensures that the correct \textit{Root Signature} is used for each instance in the \textit{Acceleration Structure}. Notice the use of method \lstinline[language=cpp]{setRootArgumentsType()}, which sets the struct type of the data in the \textit{Root Signature}. The actual values for those types are provided by the \textit{Shader Table} entries, fetched at traversal time. Those entries contain the indexing data for the instance buffer in the \textit{Global Root Signature}.

\begin{lstlisting}[caption={Local Root Signatures creation.}, label={code:local_root}, language=cpp]

MandelJuliaPacSceneBuilder::buildLocalRootSignatures()
{
    // Triangle geometry local root signature.
	auto triangleSignature = make_shared<RootSignature>(
	    "Triangle", deviceResources, m_dxr->descriptorHeap, true);
	triangleSignature->addConstant(
	    RootComponent(PrimitiveConstantBuffer()), 1);

	// Root Arguments type.
	triangleSignature->setRootArgumentsType(
	    RootArguments(TriangleRootArguments()));
	m_scene->addLocalSignature(triangleSignature);

	// Procedural geometry local root signature.
	auto proceduralSignature = make_shared<RootSignature>(
	    "Procedural", deviceResources, m_dxr->descriptorHeap, true);
	proceduralSignature->addConstant(
	    RootComponent(PrimitiveConstantBuffer()), 1);
	proceduralSignature->addConstant(
	    RootComponent(PrimitiveInstanceConstantBuffer()), 2);

	// Root Arguments.
	proceduralSignature->setRootArgumentsType(
	    RootArguments(ProceduralRootArguments()));
	m_scene->addLocalSignature(proceduralSignature);
}

\end{lstlisting}

The flexibility comes from the \textit{Shader Table} entries. They are used to build both the \textit{Shader Table} and the \textit{RTPSO}. The \textit{Shader Table} is responsible for associating \textit{Hit groups} and \textit{Local Root Signature Arguments} with the \textit{Acceleration Structure}, so the indices for the instance buffer in the \textit{Global Root Signature} can be fetch in the traversal. Listing~\ref{code:shader_table} shows the process. Remember that a \textit{Shader Table} entry has the format \textit{$<$Ray id, Hit Group id, Local Root Signature id, Root Arguments$>$}. The most important here is to note how \lstinline[language=cpp]{ProceduralRootArguments rootArgs} is set to contain the indices for the instance buffer.

\pagebreak

\begin{lstlisting}[caption={Shader table entries creation.}, label={code:shader_table}, language=cpp]

void MandelJuliaPacSceneBuilder::buildShaderTablesEntries()
{
	m_shaderTable = make_shared<RtxEngine::ShaderTable>(
	    m_scene, m_dxr->deviceResources);
	m_shaderTable->addRayGen(L"Raygen");
	m_shaderTable->addMiss("Radiance");
	m_shaderTable->addMiss("Shadow");

	// Triangle Hit Groups.
	TriangleRootArguments rootArgs{ m_planeMaterialCB };
	m_shaderTable->addCommonEntry(ShaderTableEntry{
	    "Radiance", "Triangle", "Triangle", rootArgs });
	m_shaderTable->addCommonEntry(ShaderTableEntry{
	    "Shadow", "Triangle_Shadow", "Triangle", rootArgs });

	// Procedural hit groups.
	UINT primitiveIndex = 0;
	UINT instanceIndex = 0;
	UINT i = 0;
	for (auto geometry : m_scene->getGeometry())
	{
		if (geometry->getType() == Geometry::Procedural)
		{
			for (auto instance : *geometry->getInstances())
			{
				ProceduralRootArguments rootArgs;
				rootArgs.materialCb=m_aabbMaterialCB[primitiveIndex];
				rootArgs.aabbCB.primitiveType = primitiveIndex;
				rootArgs.aabbCB.instanceIndex = instanceIndex;

				string radianceHitGroup;
				string shadowHitGroup;

				switch (primitiveIndex)
				{
				case SignedDistancePrimitive::Mandelbulb:
				{
					radianceHitGroup = "Mandelbulb";
					shadowHitGroup = "Mandelbulb_Shadow";
					break;
				}
				case SignedDistancePrimitive::Pacman:
				{
					radianceHitGroup = "Pacman";
					shadowHitGroup = "Pacman_Shadow";
					break;
				}
				
				case SignedDistancePrimitive::JuliaSets:
				{
					radianceHitGroup = "Julia";
					shadowHitGroup = "Julia_Shadow";
					break;
				}
				}

				m_shaderTable->addCommonEntry(ShaderTableEntry{
				    "Radiance", radianceHitGroup, "Procedural",
				    rootArgs });
				m_shaderTable->addCommonEntry(ShaderTableEntry{
				    "Shadow", shadowHitGroup, "Procedural",
				    rootArgs });

				++instanceIndex;
			}
			++primitiveIndex;
		}
	}
}

\end{lstlisting}

\section{Conclusion}
\label{sec:conclusion}

This paper introduced Proceduray, an engine for real-time ray tracing of procedural geometry. We discussed in detail the problems related to integrating DXR procedural geometry in engines. We also discussed DXR host code in detail, a topic with very scarce references.

This work provides an in-depth example of how to create the host code for a real-time scene with non-trivial procedural objects, such as the Mandelbulb and Julia Sets, using Proceduray. For a full implementation with host and device code, please check the official repository at {\small \url{https://github.com/dsilvavinicius/Proceduray}}. For details about the device code, please check~\cite{dasilva2021fractals}.

We hope this paper can help developers to use DXR in the context of procedural geometry and can clarify how to develop flexible intersection shaders using \textit{Shader Tables}. Additionally, we expect it to foment a more open discussion about DXR host code.

Future works include using Proceduray in more applications involving procedural geometry. Particularly, we are interested in supporting efficient ways to represent curved rays (\cite{velho2020immersive,novello2020visualization,global_illumination,novello2020design}). We also plan to investigate support for Neural Implicits. Finally, we are considering supporting Vulkan~\cite{sellers2016vulkan} in future versions.

\small
\bibliographystyle{jcgt}
\bibliography{proceduray}

\afterdoc

\end{document}